\documentclass{iopart}
\usepackage{iopams,psfig}
\begin{document}
\title{Quantum tomography as a tool for the characterization of optical
devices}
\author{G. Mauro D'Ariano ${}^{\star}$, Martina De
Laurentis${}^{\dagger}$,
Matteo G. A. Paris${}^{\star}$, Alberto Porzio${}^{\dagger \ddagger}$,
Salvatore Solimeno${}^{\dagger \ddagger}$}
\address{${}^{\star}$
Quantum Optics and Information Group, Unit\`a INFM and Dipartimento di
Fisica "Alessandro Volta", Universit\`a di Pavia, via Bassi 6, I-27100
Pavia, Italia}
\address{${}^{\ddagger}$ INFM Unit\'a di Napoli}
\address{${}^{\dagger}$ Dipartimento di Scienze Fisiche,
Universit\`a "Federico II", Complesso Universitario di Monte Sant'Angelo,
via Cintia, 80126 Napoli, Italy}
\begin{abstract}
We describe a novel tool for the quantum characterization of optical devices.
The experimental setup involves a stable reference state that undergoes an
unknown quantum transformation and is then revealed by balanced homodyne
detection. Through tomographic analysis on the homodyne data we are able to
characterize the signal and to estimate parameters of the interaction, such as
the loss of an optical component, or the gain of an amplifier. We present
experimental results for coherent signals, with application to the estimation
of losses introduced by simple optical components, and show how these results
can be extended to the characterization of more general optical devices.
\end{abstract}
\section{Introduction}
Quantum homodyne tomography (QHT) is certainly the most successful
technique for measuring the quantum state of radiation.  It is based
on homodyne detection, where the signal mode is amplified by the local
oscillator. This means that there is no need for single-photon
resolving photodetectors, whence it is possible to achieve quantum
efficiency $\eta$ approaching the ideal unit value by using linear
photodiodes \cite{kum}. Moreover, QHT is efficient and 
statistically reliable, such that it can be used on-line with the experiment.
Indeed, among other proposed state reconstruction methods, QHT is the only one
which has been implemented in quantum optical experiments \cite{mca,sch}.
\par
Possible applications of QHT range from the measurement of photon correlations
on a sub-picosecond time-scale \cite{mca} to the characterization of squeezing
properties \cite{sch,fio}, photon statistics in parametric fluorescence
\cite{cbs}, quantum correlations in down-conversion \cite{kum} and
nonclassicality of states \cite{ncl}. In general, the key point is that QHT
provides information about the quantum state within a chosen sideband, thus
allowing for a precise spectral characterization of the light beam under
investigation.
\par
In this paper we address QHT as a tool for the quantum characterization of
optical devices, like the estimation of the coupling constant of an active
medium or the quantum efficiency of a photodetector.  The goal is to link the
estimation of such parameters with the results from feasible measurement
schemes, as homodyne detection, and to make the estimation procedure the most
efficient. We present our experimental results about the reconstruction of the
quantum state of coherent signals, together with application to the estimation 
of the losses introduced by simple optical components. Moreover, we show how 
these preliminary results can be extended to the characterization of
more general optical devices.
\par
In the next Section we review some basic elements of quantum tomography, 
whereas in Section \ref{s:tool} we describe the basic requirements needed 
to implement a quantum characterization tool based on tomographic
measurements. In Section \ref{s:exp} the experimental apparatus is described 
with some details, and in Section \ref{s:data} the experimental data are 
analyzed and discussed. Section \ref{s:outro} closes the paper by discussing 
the possible extensions of the present work. 
\section{Quantum Homodyne Tomography}
Quantum tomography of a single-mode radiation field consists of  a set of
repeated measurements of the field-quadrature $x_{\phi}=\frac{1}{2}
(ae^{-i\phi}+a^{\dag}e^{i\phi})$ at different values of the reference phase
$\phi$. The expectation value of a generic operator can be obtained by
averaging a suitable kernel function $R[O](x,\phi)$ as follows \cite{lx0}
\begin{eqnarray}
\hbox{Tr}\left\{\varrho\:O\right\}=
\int_0^{\pi}\!\!\frac{d\phi}{\pi} \int_{-\infty}^{\infty} \!\!
dx\;p(x,\phi)\;R[O](x,\phi)\label{avprob}\;,
\end{eqnarray}
where $p(x,\phi)$ denotes the probability distribution of the
outcomes
$x$ for the quadrature $x_{\phi}$, and
$R[O](x,\phi)$ is given by
\begin{eqnarray}
R[O](x,\phi) = \frac{1}{4}\int_{0}^{\infty} \!\! dr
\: \hbox{Tr}\left\{O \: \cos \left[\sqrt{r} (x-{x}_{\phi})\right] \right\}
\label{kerdef}\;.
\end{eqnarray}
Actually, the tomographic kernel $R[O](x,\phi)$ for a given operator $O$ is
not unique, since a large class of {\em null functions} \cite{dak,obs}
$F(x,\phi)$ exists that have zero tomographic average for arbitrary state.
This degree of freedom can be exploited to {\em adapt} the kernel to data and
achieve an optimized determination of the quantity of interest. For example, 
quantities like the photon number, the field amplitude and any normally 
ordered moment can be the tomographically estimated by averaging the following 
kernel
\begin{eqnarray}
K[O](x,\phi)= R[O](x,\phi) + \sum_{k=0}^{M-1} \mu_k F_k(x,\phi)
+ \sum_{k=0}^{M-1} \mu_k^* F_k^*(x,\phi) \label{knull1}\;,
\end{eqnarray}
where the kernels $R[O]$ for the moments are given by \cite{ric}
\begin{eqnarray}
R[a^{\dag}{}^n a^m](x;\phi)=e^{i(m-n)\phi}
\frac{H_{n+m}(\sqrt{2}\:x)}{\sqrt{2^{n+m}}{{n+m}\choose n}}\label{ric}\;,
\end{eqnarray}
$H_n(x)$ being the Hermite polynomial of order $n$, and the {\em null
functions} $F_k$ are expressed as $F_k(x,\phi) = x^k e^{i(k+2)\phi}\:,\:
k=0,1,...$ The coefficients $\mu_k$ are obtained by minimizing the rms error
for the given kernel on the given homodyne sample. A similar approach can be
applied to optimize the reconstruction of the matrix elements
$\varrho_{mn}=\langle m|\varrho|n\rangle$, thus achieving an effective quantum
state characterization.
\section{QHT as a tool to estimate parameters}\label{s:tool} 
The state reconstruction method provided by QHT is effective and reliable, 
such that it can be exploited to build a tool to characterize quantum devices.
The general scheme for such a tool should be as follows. First, we need a
stable source of quantum states, {\em i.e.} a source able to provide repeated
preparations of a reference signal. The signal can be characterized by QHT,
and then employed as input of a given device, which we want to characterize by
the estimation of some relevant parameters. At the output, the transformed
state can analyzed by QHT, such to characterize the input-output relations of
the device.
\par
In order to implement this kind of scheme two basic requirements should be 
satisfied: i) we need a stable source for the reference signal, and ii) an
effective data processing for the tomographic samples should be devised, in
order to minimize the number of measureements. 
\par
The first point can be satisfied by considering Gaussian signals, like
coherent or squeezed states.  Indeed, quantum signals that are most likely to
be reliably generated in a lab are Gaussian states. The most general Gaussian
state can be written as $\varrho =D(\mu)\,S(r)\, \nu \, S^\dag
(r)\,D^\dag (\mu)\;, \label{xx}$ where $\nu$ denotes a thermal state
$\nu=(n_{th}+1)^{-1}[n_{th}/(n_{th}+1)]^{a^\dag a}$, $S(r)=\exp[r(a^2-a^{\dag
2})/2]$ the squeezing operator and $D(\mu)=\exp(\mu a^\dag -\mu^*a)$ the
displacement operator. However, thermal excitations can be neglected at
optical frequencies, such that we may generally consider $\nu$ as the vacuum
state.  The homodyne distribution of the state $\varrho$ at phase $\phi$ with
respect to the local oscillator is Gaussian and, remarkably, such Gaussian
character is not altered by many transformations induced by optical devices,
such as the loss of a component, the gain of an amplifier or the quantum
efficiency of a detector. In this paper we consider the reference signal
excited in a coherent state.  More general signals will be considered
elsewhere.
\par
The need of an effective data processing lead to consider either adaptive or
maximum-likelihood (ML) procedures on the tomographic data.  In Section
\ref{s:data} we apply adaptive tomography for the estimation of losses induced
by optical filters. Here, we illustrate the use of ML procedure to the
characterization of a general (active or passive) optical media, which we plan
to perform experimentally in the near future.  
\par
Let us start by reviewing the ML approach. Let $p(x | \lambda)$ the
probability density of a random variable $x$, conditioned to the value of the
parameter $\lambda$.  The analytical form of $p$ is known, but the true value
of the parameter $\lambda$ is unknown, and should be estimated from the result
of a measurement of $x$.   Let $x_1, x_2, ..., x_N$ be a
random sample of size $N$. The joint probability density of the independent
random variable $x_1, x_2, ..., x_N$ (the global probability of the sample) is
given by
\begin{eqnarray}
{\cal L}(\lambda)= \Pi_{k=1}^N \: p(x_k |\lambda)
\label{likdef}\;,
\end{eqnarray}
and is called the likelihood function of the given data sample.  The ML
estimator  of the parameter $\lambda$ is defined as the quantity $\lambda_{ML}$ 
that maximizes ${\cal L} (\lambda)$ for variations of $\lambda$.  Since
the likelihood is positive this is equivalent to maximize
\begin{eqnarray}
L(\lambda) = \log {\cal L} (\lambda) = \sum_{k=1}^N \log p(x_k | \lambda)
\label{loglikfun}\;
\end{eqnarray}
which is the so-called log-likelihood function.
\par
Let us consider a generic optical media: 
the propagation of a signal is governed, for
negligible saturation effects, by the master equation
\begin{eqnarray}
\dot \varrho =G_1\, L[a]\,\varrho +G_2\, L[a^\dag ]\,\varrho
\;,\label{meq}
\end{eqnarray}
where $\varrho$ is the density matrix describing the quantum state of the
signal mode  $a$ and $L[O]$ denotes the Lindblad superoperator $L[O]
A=O\,A\,O^\dag -\frac 12 O^\dag O\,A - \frac 12 \,A\,O^\dag O$.  If we model
the propagation as the interaction of a traveling wave single-mode $a$  with a
system of N identical two-level atoms, then the absorption $G_1=\gamma N_1$
and amplification $G_2=\gamma N_2$ parameters are related to the number $N_1$
and $N_2$ of atoms in the lower and upper level respectively. The quantity
$\gamma$ is a rate of the order of the atomic linewidth \cite{mandel}, and the
propagation gain (or deamplification) is given by ${\cal G}=\exp [(G_2-G_1)\:
t]$.  A medium described by the master equation (\ref{meq}) represents
a kind of phase-insensitive optical device, such that the parameters $G_1$ and
$G_2$ can be estimated starting from random phase tomographic data.  According
to (\ref{meq}), the homodyne distribution of a coherent signal with initial
amplitude  $\alpha_0$ is given, after the propagation, by
\begin{eqnarray}
p(x;\phi )=\frac {1}{\sqrt{\pi(\delta ^2 +g^2/2)}}\exp\left\{-\frac
{1}{\delta ^2+g^2/2}\left[ x-g\hbox{Re}(\alpha _0\,e^{-i\phi })\right]^2
\right\}\;.\label{x1}
\end{eqnarray}
with $g=e^{-Qt}$, $2Q=(G_1-G_2)$, and $\delta ^2=(G_1+G_2)(1-g^2)/4Q$ (for
non-unit quantum efficiency $\eta < 1$, $\delta ^2+ g^2/2 \rightarrow \delta
^2+ g^2/2 + (1-\eta)/2\eta)$.  By ML estimation on homodyne data we may
reconstruct the parameters $G_1$ and $G_2$. The resulting method has proven
efficient by numerical simulations \cite{qcmc}, and provides a precise
determination of the absorption and amplification parameters of the master
equation using small homodyne data sample. Notice that no advantage in using
squeezed states should be expected, because of the phase-insensitive character
of the device.
\section{Experiment}\label{s:exp}
\begin{figure}[h]
\psfig{file=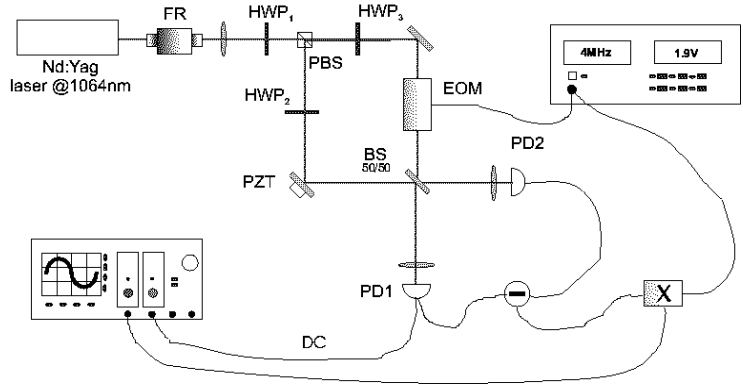,width=10cm}
\caption{Schematic of the experimental set-up. A Nd:YAG laser beam is divided
into two beam, one acts as the homodyne local oscillator, the other represent
the signal beam. The signal is modulated at frequency $\Omega$ with a defined
modulation depth to control the average photon number in the generated
coherent state. The tomographic data are provided by an homodyne detector
whose differece photocurrent is demodulated and then acquired by a digital
oscilloscope (Tektronix TDS 520D).\label{f:exp}} \end{figure}

A schematic of the experiment is presented in Fig. \ref{f:exp}. The
principal radiation source is provided by a monolithic Nd:YAG laser ($%
\approx $ 50 mW @1064 nm; Lightwave, model 142). The laser has a linewidth
of less than 10 kHz/ms with a frequency jitter of less than 300 kHz/s, while
its intensity spectrum is shot--noise limited above 2.5 MHz. The laser emits
a linearly polarized beam in a TEM00 mode.
\par
The laser is protected from back reflection by a Faraday rotator (FR) whose
output polarizing beam splitter can be adjusted to obtain an output beam of
variable intensity. The beam outing the isolator, of $\approx $ 2.5
mW, is then split into two part of variable relative intensity by a
combination of a halfwave plate (HWP$_{1}$) and a polarizing beam splitter
(PBS, p--transmitted, s--reflected). The strongest part, is directly sent toward
the homodyne beam--splitter (BS) where it acts as the local oscillator beam. One
of the mirror in the local oscillator path is piezo mounted to obtain a
variable phase difference between the two beam. The remaining part,
typically less than 200\thinspace $\mu $W, is the homodyne signal. The
optical paths traveled by the local oscillator and the signal beams are
carefully adjusted to obtain a visibility typically above $75\%$ measured at
one of the homodyne output port. The signal beam is modulated, by means of a
phase electro--optic modulator (EOM, Linos Photonics PM0202), at 4MHz, and a
halfwave plate (HWP$_{2}$, HWP$_{3}$) is mounted in each path to carefully
match the polarization state at the homodyne input.
\par
The basic property of the homodyne detector (described into details in Ref. 
\cite{ktp}) is a narrow--band detection of the field fluctuations around
4MHz. The detector is composed by a 50$\div $50 beam splitter (BS), two amplified
photodiodes (PD1, PD2), and a power combiner. The difference photocurrent is
demodulated at 4MHz by means of an electrical mixer. In this way the
detection occurs outside any technical noise and, more important, in a
spectral region where the laser does not carry excess noise.
\par
The phase modulation added to the signal beam move a certain number of
photons, proportional to the square of the modulation depth, from the
carrier optical frequency $\omega $ to the side bands at $\omega \pm \Omega $
so generating two few photons coherent state, with engineered average photon
number, at frequencies $\omega \pm \Omega $. The sum sideband modes is
then detected as a controlled perturbation attached to the signal beam \cite
{sch}. The demodulated current is acquired by a digital oscilloscope
(Tektronix TDS 520D) with 8 bit resolution and record length of 250000
points per run. The acquisition is triggered by a triangular shaped waveform
applied to the PZT mounted on the local oscillator path. The piezo ramp is
adjusted to obtain a $2\pi $ phase variation between the local oscillator
and the signal beam in an acquisition window.
\par
The homodyne data, to be used for tomographic reconstruction of the
radiation state, have been calibrated according to the noise of 
the vacuum state. This is obtained by acquiring a set of data leaving 
the signal beam undisturbed while scanning the local oscillator phase. 
It is important to note that in case of the vacuum state
no role is played by the visibility at the homodyne beam--splitter. 
\section{Data Analysis}\label{s:data}
 Our tomographic samples consist of $N$ homodyne data $%
\{x_{j},\phi _{j}\}_{j=1,...,N}$ with phases $\phi _{j}$ equally spaced with
respect to the local oscillator. Since the piezo ramp is active during the
whole acquisition time, we have a single value $x_{j}$ for any phase $\phi
_{j}$. From calibrated data we first reconstruct the quantum state of the 
homodyne signal. According to the
experimental setup described in the previous section we expect a coherent
signal with nominal amplitude that can be adjusted by varying the modulation
depth of the optical mixer. However, since we do not compensate for the
quantum efficiency of photodiodes in the homodyne detector ($\eta \simeq
90\% $) we expect to reveal coherent signals with reduced amplitude 
with respect to actual one.  In addition, the amplitude is furtherly reduced
by the non-maximum visibility (ranging from $75\%$ to $85\%$) at the
homodyne beam--splitter.
\par
\begin{figure}[h]
\begin{tabular}{lcr}
\psfig{file=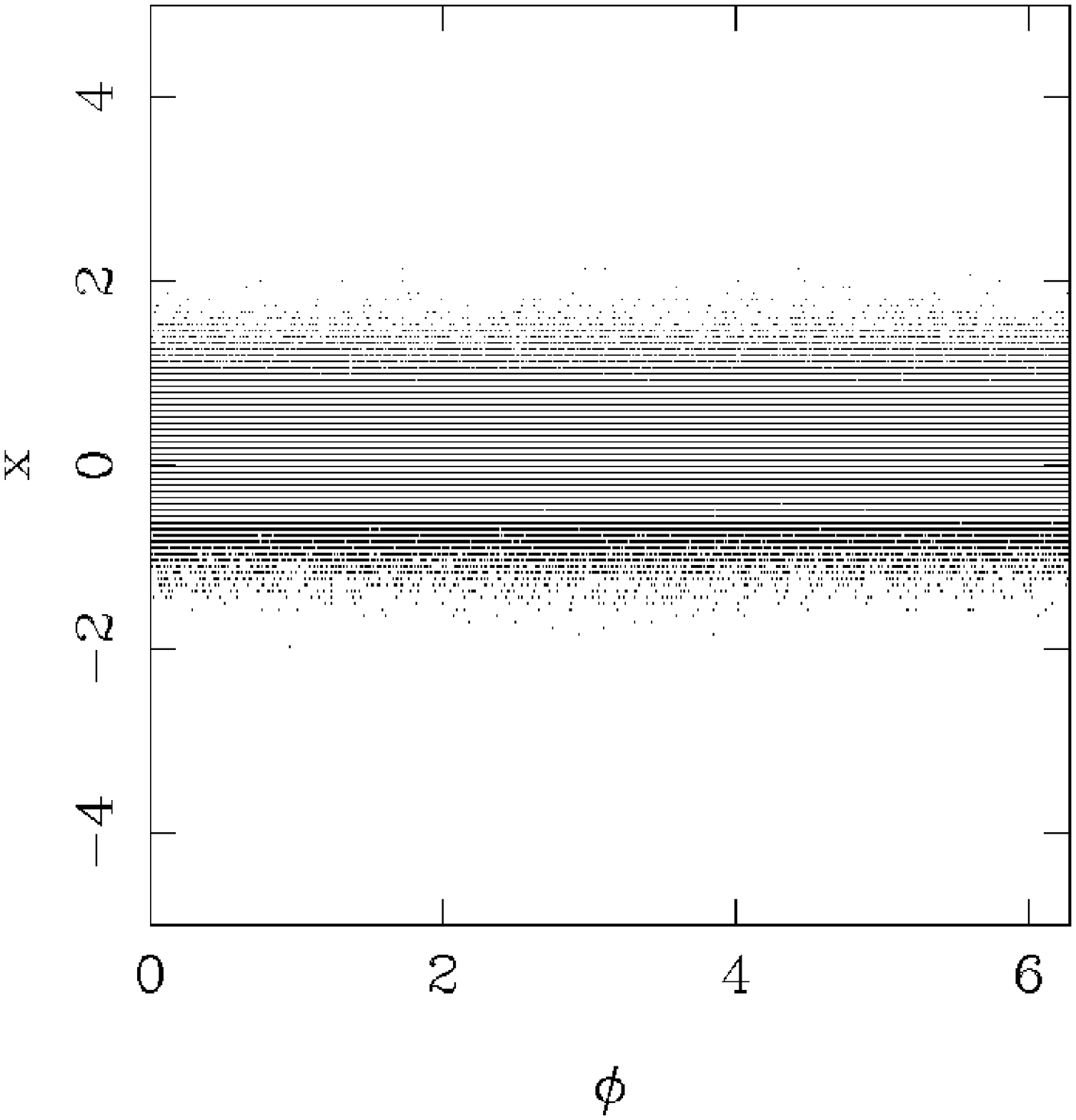,width=4cm} &
\psfig{file=20set01-75mv_2.ps,width=4cm} &
\psfig{file=20set01-75mv_3.ps,width=4cm} \\
\psfig{file=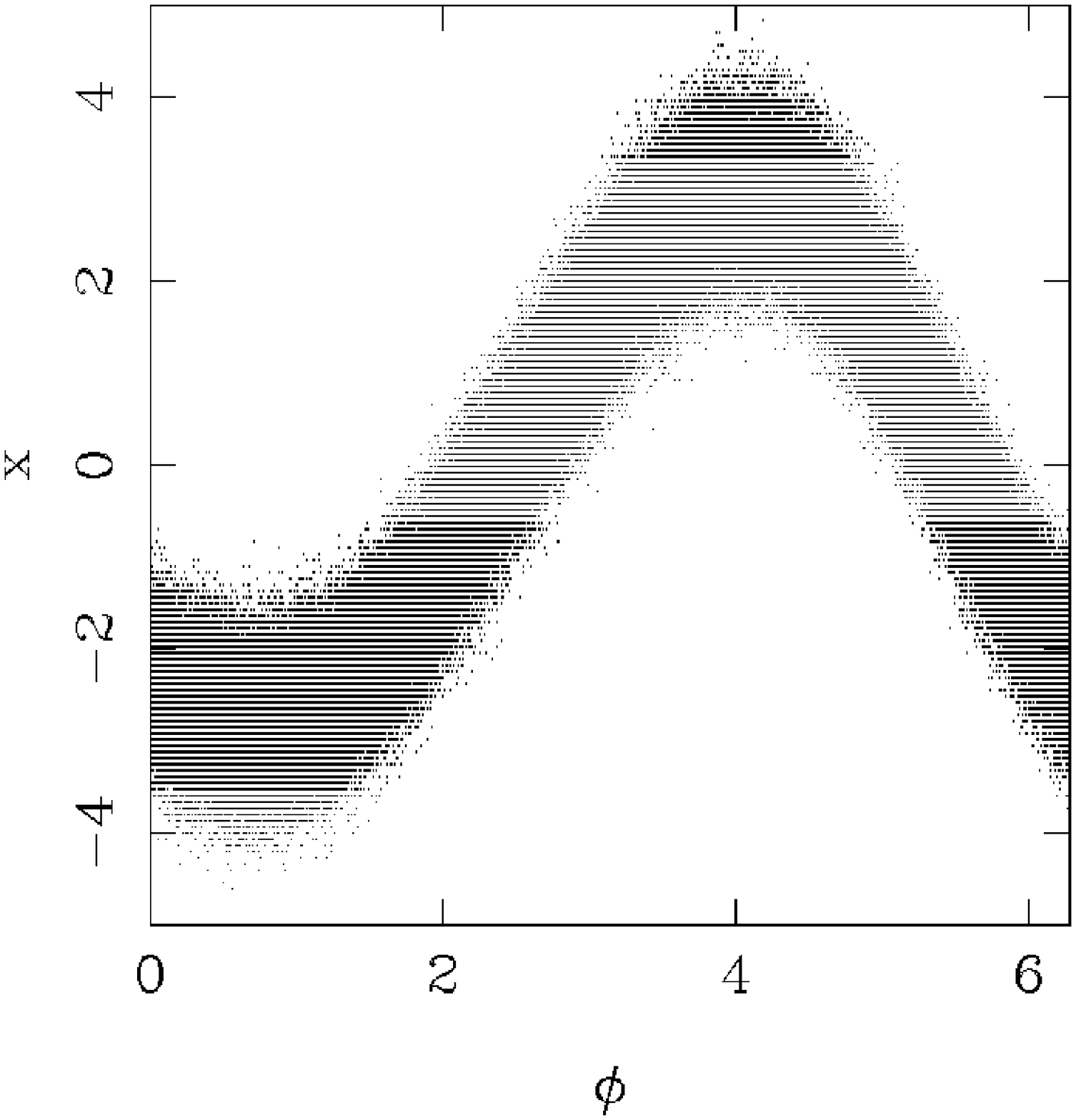,width=4cm} &
\psfig{file=20set01-75mv_5.ps,width=4cm} &
\psfig{file=20set01-75mv_6.ps,width=4cm}
\end{tabular}
\caption{Reconstruction of the quantum state of the signal, and of the vacuum 
state used for calibration. For both states, from left to right, we report the 
raw data, a histogram of the photon number distribution, and a contour plot of 
the Wigner function. Reconstruction has been performed by a sample of $N=242250$ 
homodyne data. The coherent signal has an estimated average photon number equal 
to $\langle a^\dag a\rangle=8.4$. The solid line denotes the theoretical photon 
distribution of a coherent state with that number of photons. Statistical errors
on matrix elements are about $2\%$ \label{f:rec}. The slight phase asymmetry
in the Wigner distribution corresponds to a value of about the $2\%$ of the
maximum.}
\end{figure}

In Fig. \ref{f:rec} we show a typical reconstruction, together with the
reconstruction of the vacuum state used for calibration. For both states, we
report the raw data, the photon number distribution, {\em i.e.} the diagonal
elements $\varrho_{nn}\equiv \langle n| \varrho|n\rangle$ of the density
matrix in the Fock representation, and a contour plot of the Wigner
function. The matrix elements are obtained by sampling the corresponding
kernel functions 
\[
R[\:|n\rangle\langle n+k|\:] (x,\phi) = 2 \exp\left\{-ik \phi \right\}\sqrt{%
2^k n!(n+k)!} \: f_{nk}(x)\:, 
\]
where 
\[
\fl f_{nk}(x)=\left\{ 
\begin{array}{lr}
\sum_{l=0}^n \frac{(-)^l 2^l\Gamma (1+l+k/2)}{l!(n-l)!(l+k)!} \Phi
(1+l+k/2,1/2;-2x^2) & k \:\: \hbox{even} \\ 
&  \\ 
\sum_{l=0}^n \frac{(-)^l 2^{l+1/2}\Gamma (1+l+(k+1)/2) }{l!(n-l)!(l+k)!} 2x
e^{-2x^2} \Phi (-l-k/2,3/2;2x^2) & k \:\: \hbox{odd}
\end{array}
\right.\:, 
\]
and $\Phi(a,b;x)$ denotes a confluent hypergeometric function. The
tomographic determination of the matrix elements is given by the averages 
\begin{eqnarray}
\varrho_{nk} = \overline{R[\:|n\rangle\langle k|\:]}=\frac1N \sum_j R[\:
|n\rangle\langle k|\:] (x_j,\phi_j) \;,  \label{rhonn}
\end{eqnarray}
whereas the corresponding confidence intervals are given (for diagonal
elements) by $\delta\varrho_{nn}=\Delta\varrho/\sqrt{N}$, $\Delta\varrho$
being the rms deviation of the kernel $R$ over data (for off-diagonal
elements the confidence intervals are evaluated for the real and imaginary
part separately).
\par
In order to see the quantum state as a whole, we also report the
reconstruction of the Wigner function of the field, which is defined as
follows 
\begin{eqnarray}
W (z) = \frac{2}{\pi} \hbox{Tr} \left\{ \varrho \: D(2z)\: (-)^{a^\dag a}
\right\} \;,  \label{wdef}
\end{eqnarray}
and can be expressed in terms of the matrix elements as 
\begin{eqnarray}
W(z)=\mbox{Re} \sum_{d=0}^{\infty} e^{id\phi} \sum_{n=0}^{\infty} \Lambda
(n,d;|z|^2) \rho_{n,n+d} \;  \label{w2}
\end{eqnarray}
where 
\begin{eqnarray}
\Lambda (n,d;|z|^2) = (-)^n 2 (2-\delta_{d0}) |2z|^d \sqrt{\frac{n!}{(n+d)!}}
e^{-2|z|^2} L_n^d (|2z|^2) \;,  \label{w3}
\end{eqnarray}
and $L_n^d (x)$ denotes the Laguerre polynomials. Of course, the series in
Eq. (\ref{w2}) should be truncated at some point, and therefore the Wigner
function can be reconstructed only at some finite resolution. 
\par
Once the coherence of the signal has been established we may use QHT to
estimate the loss imposed by a passive optical component like an optical
filter. The procedure may be outlined as follows. We first estimate the
initial mean photon number $\bar{n}_{0}=|\alpha _{0}|^{2}$ of the signal
beam, and then the same quantity inserting an optical neutral density 
filter in the signal path. If $\Gamma $ is the loss parameter, then the 
coherent amplitude is reduced to $\alpha _{\Gamma }=\alpha _{0}e^{-\Gamma }$, 
and the intensity to $\bar{n}_{\Gamma }=\bar{n}_{0}e^{-2\Gamma }$.
\par
The estimation of the mean photon number can be performed adaptively on data
by taking the average of the kernel 
\begin{eqnarray}
K[a^{\dag} a](x)=2 x^2 - \frac{1}{2} + \mu e^{i2\phi} + \mu^* e^{-i2\phi} \;,
\label{num}
\end{eqnarray}
where $\mu$ is a parameter to be determined in order to minimize
fluctuations. As proved in Ref. \cite{obs} $\mu=-1/2 \langle a^{\dag
2}\rangle$, which itself can be obtained from homodyne data. In practice,
one uses the data sample twice: first to evaluate $\mu$, then to
obtain the estimate for the mean photon number.
\par
\begin{figure}[h]
\begin{minipage}{4cm}
\psfig{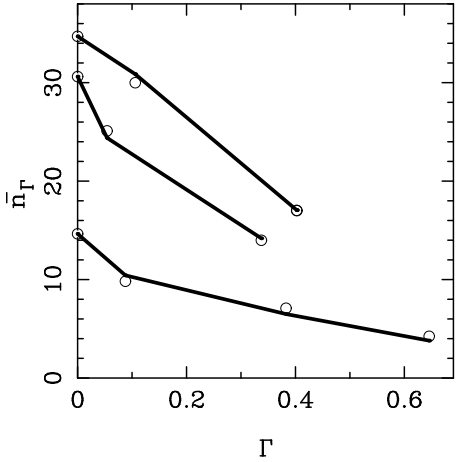}
\end{minipage}
\begin{minipage}{9cm}
\caption{Estimation of the mean photon number of a coherent signal as a
function of the loss imposed by an optical filter. Three set of experiments,
corresponding to three different initial amplitudes are reported. Open circles 
are the tomographic determinations, whereas solid line
denotes the expected values, as follow from nominal values of loss and
visibility at the homodyne. Statistical errors are within
the circles.  \label{f:loss}} \end{minipage}
\end{figure}

In Fig. \ref{f:loss} the tomographic determinations of $\bar{n}_\Gamma$ 
are compared with the expected values for three set of experiments, corresponding 
to three different initial amplitudes. The expected values are given by 
$\bar{n}_{\Gamma }=\bar{n}_{0}e^{-2\Gamma }{\cal V}$, where $\Gamma $ is the value 
obtained by comparing the signal dc currents $I_{0}$ and $I_{\Gamma }$ at
the homodyne photodiodes and ${\cal V=V}_{\Gamma }/{\cal V}_{0}$ is the 
relative visibility. The
solid line in Fig. \ref{f:loss} denotes these values. The line is not continuos
due to variations of visibility. As it is apparent from the plot the
estimation is reliable in the whole range of values we could explore. It is
worth noticing that the present estimation is absolute, {\em i.e.} it does
not depends on the knowledge of the initial amplitude, and it is robust,
since it may performed independently on the quantum efficiency of the
photodiodes employed for the homodyne detector. 
\section{Conclusions and outlook}\label{s:outro}
In this paper we addressed QHT as a tool for the characterization of 
quantum optical devices. We carried out the quantum state reconstruction of
coherent signals, and show how QHT can be used to reliably estimate the loss 
imposed by an optical filter. We also show that the estimation procedure can 
be extended to the characterization of general (active or passive) optical 
devices, which we plan to perform experimentally in the near future. We also 
plan to extend our analysis to squeezed signals since it has been proved 
\cite{parlik} that squeezing improves precision in the homodyne estimation 
of relevant parameters such the phase-shift or the quantum efficiency of a 
photodetector. In particular, our aim is to reproduce the described measurement 
scheme at the output of an OPO cavity \cite{lnb} driven below threshold to 
generate vacuum squeezed radiation.
\section*{Acknowledgment}
This work has been sponsored by INFM under the project PAIS TWIN. 

\end{document}